\documentclass[12pt, one-column, superscript address]{revtex4-2}
\usepackage[a4paper,bindingoffset=0.2in,%
           left = 1.52cm, right = 1.52cm, top = 3.1cm, bottom = 2.4cm,%
            footskip=.25in]{geometry}
\usepackage{graphicx} 
\usepackage{tcolorbox}
\usepackage{todonotes}

\usepackage{subfigure}
\usepackage{float}
\usepackage{comment}
\usepackage{titlesec}
\usepackage{hyperref, bookmark}
\usepackage{tcolorbox}
\tcbuselibrary{skins,breakable}
\usepackage{color}
\usepackage[version=4]{mhchem}
\usepackage{siunitx}
\usepackage{soul}
\usepackage[ruled,linesnumbered]{algorithm2e}

\usepackage{ mathrsfs }
\usepackage{ braket }
\usepackage{ mathtools, slashed }
\usepackage{bbold}
\usepackage{amsmath}

\DeclareMathOperator*{\argmin}{arg\,min}

\usepackage{hyperref, bookmark}
\bookmarksetup{
  numbered, 
  open,
}

\newcommand{\at}[1]{{\color{blue}[AT: #1]}}

\hypersetup{
  colorlinks = true
}

\begin{document}

\title{Contrasting Statistical Phase Estimation with the Variational Quantum Eigensolver in the era of Early Fault Tolerant Quantum Computation}

\author{Ming-Zhi Chung}
\email{chung@qunasys.com}
\affiliation{QunaSys Inc.,
Aqua Hakusan Building, 9th Floor, 1-13-7 Hakusan, Bunkyo-ku, Tokyo 113-0001, Japan }

\author{Andreas Thomasen}
\affiliation{QunaSys Inc.,
Aqua Hakusan Building, 9th Floor, 1-13-7 Hakusan, Bunkyo-ku, Tokyo 113-0001, Japan }

\author{Henry Liao}
\affiliation{QunaSys Inc.,
Aqua Hakusan Building, 9th Floor, 1-13-7 Hakusan, Bunkyo-ku, Tokyo 113-0001, Japan }
\affiliation{Department of Physics and Astronomy, National Taiwan University, Taipei 10617, Taiwan }

\author{Ryosuke Imai}
\affiliation{QunaSys Inc.,
Aqua Hakusan Building, 9th Floor, 1-13-7 Hakusan, Bunkyo-ku, Tokyo 113-0001, Japan }

\date{\today}

\begin{abstract}
    In this review, we give an overview of the proposed applications in the early-FTQC (EFTQC) era.
    Starting from the error correction architecture for EFTQC device, we first review the recently developed space-time efficient analogue rotation (STAR) architecture \cite{akahoshiPartiallyFaultTolerantQuantum2024}, which is a partially fault-tolerant error correction architecture.
    Then, we review the requirements of an EFTQC algorithm.
    In particular, the class of ground state energy estimation (GSEE) algorithm known as the statistical phase estimation algorithm (SPE) is studied.
    We especially cast our attention on two SPE-type algorithms, the step-function filter-based variant by Lin and Tong (LT22) \cite{Lin:2021rwb} and Gaussian Filter \cite{Wang:2022gxu}.
    Based on the latter, we introduce the Gaussian Fitting algorithm, which uses an alternative post-processing procedure compared to \cite{Wang:2022gxu}. 
    Finally, we systematically simulate the aforementioned algorithms and Variational Quantum Eigensolver (VQE) using the 1-uCJ ansatz with different shot counts.
    Most importantly, we perform noisy simulations based on the STAR architecture.
    We find that for estimating the ground state energy of the 4-qubit $\ce{H}_2$ Hamiltonian in the STO-3G basis, SPE becomes more advantageous over VQE when the physical error rate is sufficiently low.
\end{abstract}

\maketitle

\section{Introduction}
The perfect device does not exist. This has always been true in computing, whether quantum or classical. In conventional computing the need to correct bit errors is so inherent that it is handled at the hardware level — even if you wanted to, you would not be able to run a modern computer without error correction. It should therefore come as no surprise that architectural considerations are taking place in quantum computing.

Fault-tolerant quantum computing (FTQC) is quantum computing based on error-correcting codes \cite{devitt2013quantum, gottesman1998theory, paler2015introduction}. In analogy with classical computation, additional computational resources are delegated to detecting and correcting errors. However, whereas bit errors that make it past these checks are extremely rare in classical architectures and the error correction overhead is relatively modest, for quantum computers the story is much more involved.

As an example, one can consider surface codes \cite{fowler2010surface, fowler2012surface}, which define logical qubits, i.e. the ones that comprise the circuits defined by a quantum algorithm, by the interactions of a large number of physical qubits. Here it is often assumed that millions of physical qubits are required for each logical qubit in fully error-corrected architectures \cite{pelucchi2022potential, bartolucci2023fusion}. Current quantum devices are said to be in the noisy intermediate-scale quantum (NISQ) era \cite{preskill2018quantum}, where due to a lack of error correction, one would usually resort to error mitigation tools \cite{endoPracticalQuantumError2018, suzukiQuantumErrorMitigation2022, temmeErrorMitigationShortDepth2017} rather than actual error correction — here there is no distinction between logical and physical qubits.

However, there is likely going to be an intermediate period in the transition between fully error-corrected devices and NISQ. As devices are scaled up and more quantum resources can be dedicated to error correction, gradually deeper quantum circuits will be realized \cite{bultrini2023battle, koukoulekidis2023framework, otten2019accounting}. As argued in \cite{katabarwa2024early}, error correction of a quantum device is subject to a law of diminishing returns. Being able to perform useful quantum operations subject to the constraints of imperfect error correction defines a limit on the scalability of quantum circuits. In this regime, circuits are repeatedly executed \cite{wangQuantumAlgorithmGround2023, wangFasterGroundState2023, svoreFasterPhaseEstimation2013} and statistical tools are used to interpret the outcomes after several repetitions \cite{wiebeEfficientBayesianPhase2016, wanRandomizedQuantumAlgorithm2022, wangStatePreparationBoosters2022, tostaRandomizedSemiquantumMatrix2023}. This is the regime of early FTQC (EFTQC).

In this paper, we review the general trends in EFTQC and introduce some of the more well-known algorithms. We also supplement our understanding of these algorithms with our own benchmarks and identify regimes of interest where EFTQC has an upper hand compared to NISQ. This allows us to give a detailed account of the importance of these algorithms in the transitional period between NISQ and FTQC. Readers who are interested are directed to existing works in literature which we will not discuss in this review, including work on the relationship between error mitigation and partial error correction \cite{suzuki2022quantum}, EFTQC simulations of the Hubbard model \cite{campbell2021early}, work on robustness of EFTQC algorithms subject to noise and an analysis of the Clifford+T gate decomposition with a limited number of T-gates \cite{kuroiwa2023clifford+}.

This paper is organized as follows. In section \ref{sec: algo_overview}, we review the STAR architecture as well as the statistical phase estimation algorithm (SPE) along with the requirements of an EFTQC algorithm. Later in section \ref{sec: SPE simulation}, physical-noise-free simulation based on \cite{Lin:2021rwb} and \cite{Wang:2022gxu} are performed. We will numerically study the algorithm complexities as well as the effect caused by Trotter error. Finally in section \ref{sec: compare with VQE}, we compare SPE against Variational Quantum Eigensolver (VQE) and perform noisy simulation based on the STAR architecture noise model. The performance of the algorithm under different physical error rates is studied. Lastly in section \ref{sec: prospect}, we identify various promising directions that can be used to facilitate EFTQC algorithm research and challenges to be tackled.

\section{Algorithms designed for early fault tolerance}\label{sec: algo_overview}
In this section, we first discuss the features an EFTQC architecture should possess, which leads to the introduction of the STAR architecture \cite{akahoshiPartiallyFaultTolerantQuantum2024}. Based on the architecture discussions, we turn our attention to the ground state energy estimation (GSEE) algorithms suitable for execution on EFTQC devices. This involves an overview of currently well-known algorithms, e.g. VQE and QPE, and how they compare to the recently developed EFTQC GSEE algorithm, known as the statistical phase estimation (SPE). We will further review the construction of SPE and discuss the commonly-used performance indicators for phase-estimation-like GSEE algorithms.

\subsection{EFTQC architecture}
We have alluded to the need for partial error correction in EFTQC. Here we will briefly describe what properties an error-correcting architecture concretely should possess. We will then go into some detail with the STAR architecture, which is a partial error correction architecture developed specifically for EFTQC.

Firstly, we will point out the practical challenges associated with fully fault-tolerant architectures. The difficulty mainly arises from the implementation of error-corrected non-Clifford gates \cite{zhouMethodologyQuantumLogic2000}. Exponential quantum speed-up is only realizable when circuits consist of both Clifford and non-Clifford gates as sampling in the computational basis of quantum states prepared using only the Clifford gate set can be simulated in polynomial time on a classical computer \cite{gottesman1998heisenberg}. Non-Clifford gates, which include continuous rotations on single-qubit, are conventionally implemented as a combination of several T-gates and Clifford gates via the Solovay-Kitaev decomposition \cite{Nielsen_Chuang_2010}. The T-gates themselves are implemented using magic state distillation \cite{bravyiMagicstateDistillationLow2012, liMagicStateFidelity2015}, a costly procedure which in most realistic scenarios will demand most of the quantum resources of the physical device by orders of magnitude \cite{kim2022fault, rubin2023fault}.

Non-Clifford gates are difficult to implement in a fault-tolerant manner then, due to the complexity of the magic state distillation protocol. However, due to the trends that we are already witnessing in contemporary quantum hardware, we can make some assumptions about various sources of error on hardware and based on them make inspired choices regarding which ones to prioritize. In a nutshell, that is the approach of partial error correction.

Contemporary quantum devices exhibit analog rotation gate inaccuracies that are typically one or two orders of magnitude lower than the two-qubit Clifford gates used to entangle qubits \cite{IBMQ}. Indeed, the two-qubit gates have a tendency to have higher error rates in their physical implementations. On the other hand, the resources needed to fault-tolerantly implement single-qubit analog rotations tend to be orders of magnitude higher due to the need for a T-gate decomposition with magic state distillation.

\begin{figure}
    \centering
    \includegraphics[width=0.8\linewidth]{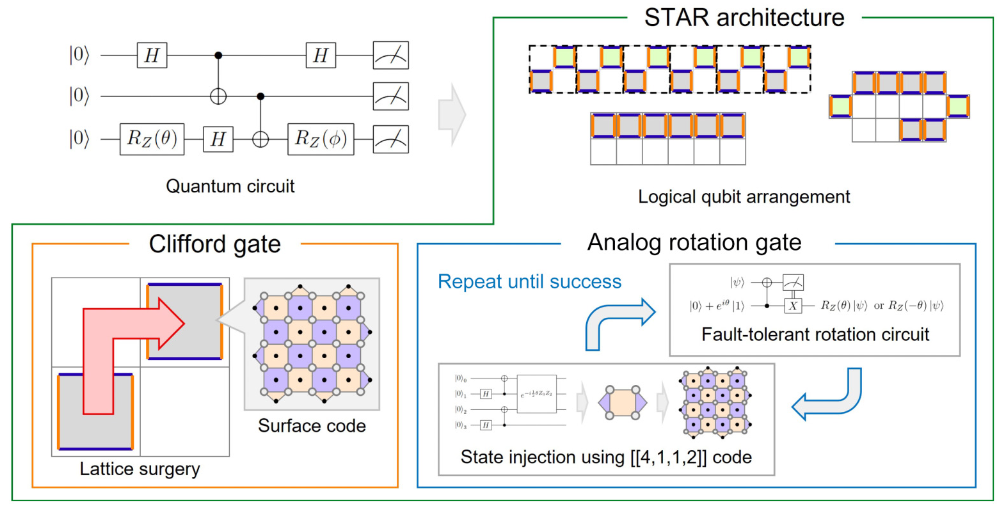}
    \caption{Schematic depiction of the STAR architecture, reproduced from \cite{akahoshiPartiallyFaultTolerantQuantum2024}. Logical qubits are made up of surface patches and Clifford gates are implemented through the rotated surface code. Meanwhile non-Clifford gates are implemented using native gates prepared through a measurement based protocol with a post-selected ancilla state that has been expanded to cover a full patch.}
    \label{fig:circuit_star_arch}
\end{figure}

One reasonable choice for EFTQC is therefore to reduce the error rate of the Clifford set using surface code error correction \cite{litinskiGameSurfaceCodes2019}, while leaving the analog rotation gates to be implemented through the native gate set. As an example, we can consider the STAR architecture~\cite{akahoshiPartiallyFaultTolerantQuantum2024}. Here the logical qubits are defined using surface patches and Clifford gates are implemented through the rotated surface code \cite{litinskiLatticeSurgeryTwist2018} using lattice surgery \cite{horsmanSurfaceCodeQuantum2012}. Analog rotations are then implemented through gate teleportation, where a single ancilla patch is used for the measurement protocol. This logical ancilla qubit has been produced by the application of native rotation gates, not by the usual distillation protocol. It can therefore implement continuous analog rotations and is not limited to only T-gates. It is encoded in the $[[4, 1, 1, 2]]$ error-detecting code and expanded to cover an entire patch with the same code distance as the rest of the patches. Having obtained this ancilla patch, an entangling operation and measurement are performed to teleport the input state into one that has undergone the specified gate rotation, consistent with the measurement-based quantum computation protocol \cite{gavrielTransversalInjectionMethod2022, briegel2009measurement}. The procedure is summarized on Figure~\ref{fig:circuit_star_arch}.

\begin{figure}
    \centering
    \includegraphics[width=0.65\linewidth]{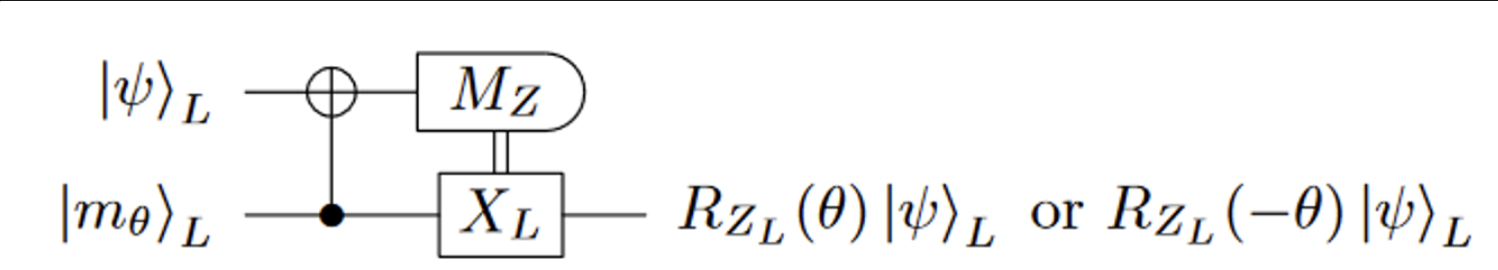}
    \caption{Gate teleportation is used to implement analog rotations in the manner shown here. This figure is reproduced from \cite{akahoshiPartiallyFaultTolerantQuantum2024}.}
    \label{fig:analog_rotation}
\end{figure}

On Fig.~\ref{fig:analog_rotation}, the logical state of the ancilla qubit is 
\begin{equation}
    \ket{m_\theta}_L = R_{Z_L}(\theta)\ket{+}_L
\end{equation}
We use the $L$ subscript to indicate that this is a logical qubit spanning several physical qubits commensurate with the chosen code-distance of the rotated surface code. We also would like to remark that we use the following convention for Pauli rotation gates
\begin{equation}
R_P(\theta) = \exp( - i \theta /2 P),
\end{equation}
Depending on the measurement outcome, the gate either performed the requested analog rotation e.g. $R_{Z_L}(\theta)$ or the unwanted rotation $R_{Z_L}(-\theta)$. By application of the same operation with $R_{Z_L}(2\theta)$ a failed rotation can be corrected. This procedure is repeated until we get the desired angle, which results in two repetitions on average. In addition, before expanding to a full patch, some post-selection is performed based on the syndromes obtained by the $[[4, 1, 1, 2]]$ code, thus providing some purification of the ancilla state. For EFTQC architecture, it is therefore reasonable to assume that

\begin{itemize}
    \item Clifford gates can be implemented in a fault-tolerant way given enough physical qubits
    \item Non-Clifford gates can be implemented with some error correction, but the error rate fundamentally depends on the fidelity of the native gate set.
\end{itemize}

\noindent Later in this article we will show the result from noisy simulations given the above conditions and some simplifying assumptions. In our noise model, Clifford gates are implemented as ideal gates without noise, whereas we assume a constant error rate for non-Clifford gates.

\begin{tcolorbox}[title=STAR architecture noise model considered in this article.]
\label{sec:star_architecture}
The native gate set under consideration is $\{\text{CNOT}, \text{H}, R_z\}$, where the Clifford gates $\{\text{CNOT}, \text{H}\}$ are error-free and the major source of error comes from logical $R_z$ gates. The post-selection performed under the $[[4,1,1,2]]$ code reduces the effective logical error rate of single analog rotations to $P_L = 2p_{\text{phys}}/15$, $p_{\text{phys}}$ being the physical error rate (see Section~VI.A of \cite{akahoshiPartiallyFaultTolerantQuantum2024}).
However, as it takes 2 attempts on average to reach the correct rotation angle, we utilize a simplified noise model where the logical error rate $P_L$ of the $R_z$ gate is described by
\begin{equation}
    P_L = 1-\left(1-\frac{2p_{\text{phys}}}{15}\right)^2 \approx \frac{4p_{\text{phys}}}{15}.
\end{equation}
Although this is a greatly simplified noise model, it captures the leading order contribution to the noise expected from transpilation to the $\{\text{CNOT}, \text{H}, R_z\}$ set. As we do not have active error correction, the dependence on the physical error rate is still $P_L\sim \mathcal{O}(p_\textup{phys})$.
\end{tcolorbox}

\subsection{Comparing VQE, QPE and SPE}\label{subsec: algo comparison}
\label{sec:nisq_eftqc_algo_comparison}

In the NISQ era, VQE is one of the most representative algorithms to solve the ground state energy estimation problem.
However, GSEE with an optimization based construction cannot be assumed to be efficiently solvable, as there are instances of this classical optimization problem that are NP-hard \cite{PhysRevLett.127.120502}, and we do not know a-priori whether a GSEE problem falls into this category or not. Another problem is that VQE is inherently a variational method, which means that Ansatz is determined heuristically. In addition, even in noiseless machines, it requires $\mathcal{O}(\epsilon^{-2})$ shots to achieve $\epsilon$ accuracy.

On the other hand, quantum phase estimation (QPE) is a powerful algorithm for estimating the ground state energy. 
Depending on the input state $|\psi_{in}\rangle$’s overlap with the exact ground state $|\lambda_0\rangle$, $p_0=|\langle \psi_\text{in}|\lambda_0\rangle|^2$, the phase estimation algorithm requires $\mathcal{O}(p_0^{-1})$ access to the phase estimation circuit and the energy accuracy achievable depends on the number of ancilla qubits, $n$, as $\mathcal{O}(2^{-n})$.
However, executing QPE at scale requires a fully fault-tolerant quantum architecture.
As we have already discussed, in the EFTQC era, error correction is performed partially in an effort to extend the maximal circuit depth achievable subject to a minimum circuit fidelity requirement.
However, in the context of QPE, it is unlikely that the number of logical qubits available will be sufficient to achieve chemical accuracy of energy error below $1.6 \text{mHa}$.
Additionally, since partial error correction will not be able to eliminate errors from non-Clifford gates, the number of available rotating gates is limited. It is not possible to sustain high-fidelity, deep, controlled time-evolution circuit execution. Given these limitations, algorithms which use quantum circuits with fewer ancilla qubits and shorter circuits are needed. 
The remaining challenge is then to restore the energy estimation accuracy through means of repeated circuit execution and statistical tools. 

The discussion above mirrors to some extent that given in \cite{Lin:2021rwb}, where a list of requirements is arrived at for energy estimation algorithms based on time evolution appropriate for EFTQC. With slight modification, we reiterate this list as

\begin{enumerate}
    \item Low circuit depth.
    \item Heisenberg-limited total run time scaling.
    \item Algorithms that show quantum advantage when limited to hundreds of logical qubits.
    \item No requirement of the exact ground state as a trial state.
\end{enumerate}
The Heisenberg limit of run time scaling is achieved when the total circuit run time dependence on the energy accuracy is $\widetilde{\mathcal{O}}(\epsilon^{-1})$, where $\widetilde{\mathcal{O}}$ corresponds to ignoring  $\text{polylog}$ factors.

A plethora of algorithms \cite{Lin:2021rwb, Wang:2022gxu, Ding:2024qvu, Kiss:2024sep, Wan:2021non, Blunt:2023gqs, Ding:2022xue, Wang:2023ruq, Wang:2022qjk} have already been developed attempting to satisfy these conditions. Among these algorithms, a class of algorithms is often referred to as Statistical Phase Estimation (SPE), which we review in the section \ref{subsec: SPE review}.
They have the potential to achieve:

\begin{itemize}
    \item Chemical accuracy with shorter circuit depth than QPE circuit.
    \item Better sampling scaling compared to VQE.
    \item Less intensive classical assists, i.e. without the need of multi-variable optimization sub-procedure.
    \item Noise resilience: When the quantum error rate is low enough, it will manifest as noise in the resulting signal. If there exists an appropriate signal processing technique, the energy eigenvalue read off from the signal can fall into the desired accuracy.
\end{itemize}

There is a trade-off between circuit depth and chemical accuracy in certain variations of SPE and in general SPE is still limited by the maximal depth achievable by the quantum architecture used. While the signal processing of SPE makes it noise resilient, this resilience disappears when the excitation gap is small or the input state is not well-prepared. The interplay of too much noise and poor state preparation can render the characteristic peaks in the signal indistinguishable. We will elaborate on these points in the next section and fully illustrate them with our benchmarking results later in Section~\ref{subsubsec: varying_noise}.

\subsection{A Lightning review of statistical phase estimation}\label{subsec: SPE review}
\begin{figure}
    \centering
    \includegraphics[width=0.5\linewidth]{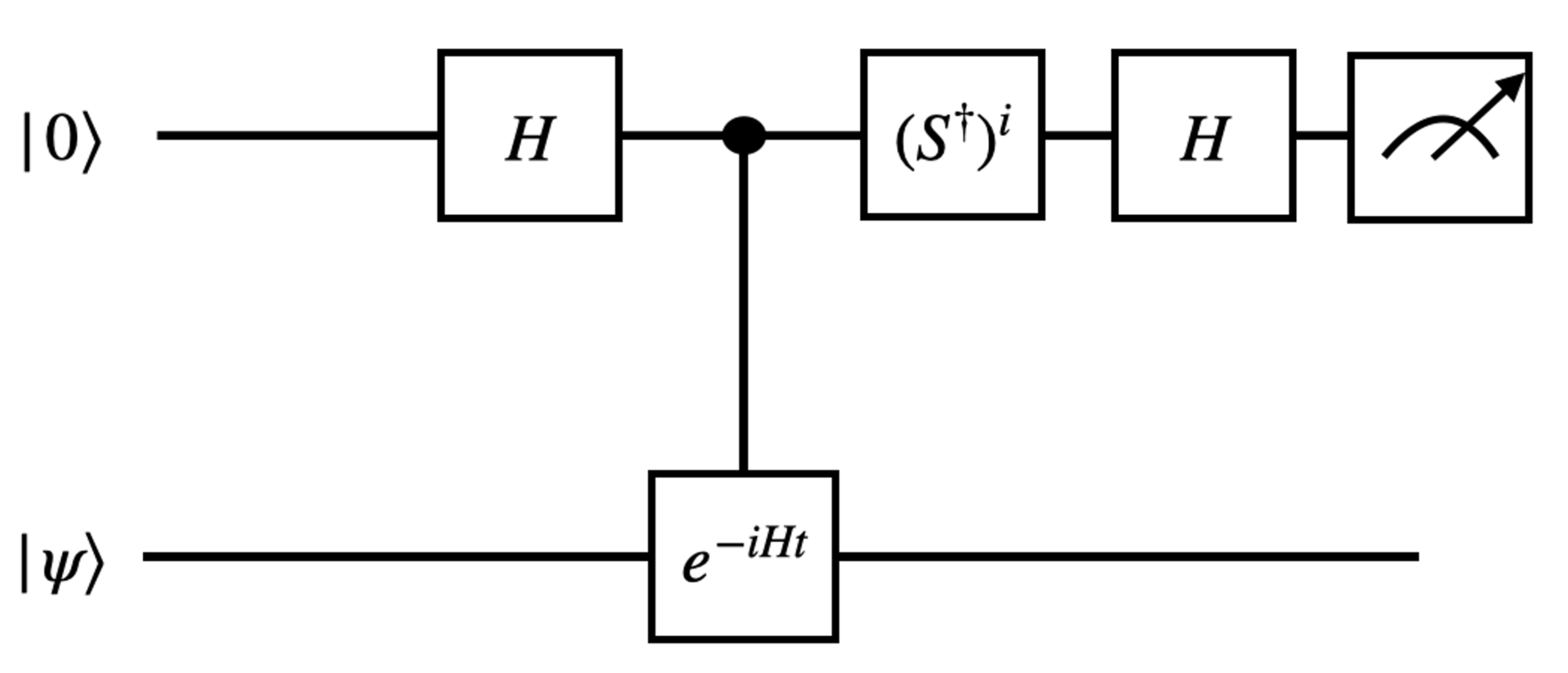}
    \caption{The Hadamard test circuit for estimating $\langle e^{-iHt}\rangle$. $i=0$ corresponds to estimating $\text{Re}[\langle e^{-iHt}\rangle]$ and $i=1$ corresponds to estimating $\text{Im}[\langle e^{-iHt}\rangle]$}
    \label{fig:hadamard_test}
\end{figure}
This section reviews the standard construction of the statistical phase estimation algorithm. This construction is widely utilized in various literatures with different variants.
In this article, we investigate two major types of statistical phase estimation: LT22 \cite{Lin:2021rwb} and Gaussian Filter \cite{Wang:2022gxu}.
Below we present the commonly shared steps among statistical phase estimation algorithms along with its derivation.



\begin{figure}
    \centering
    \includegraphics[width=0.95\linewidth]{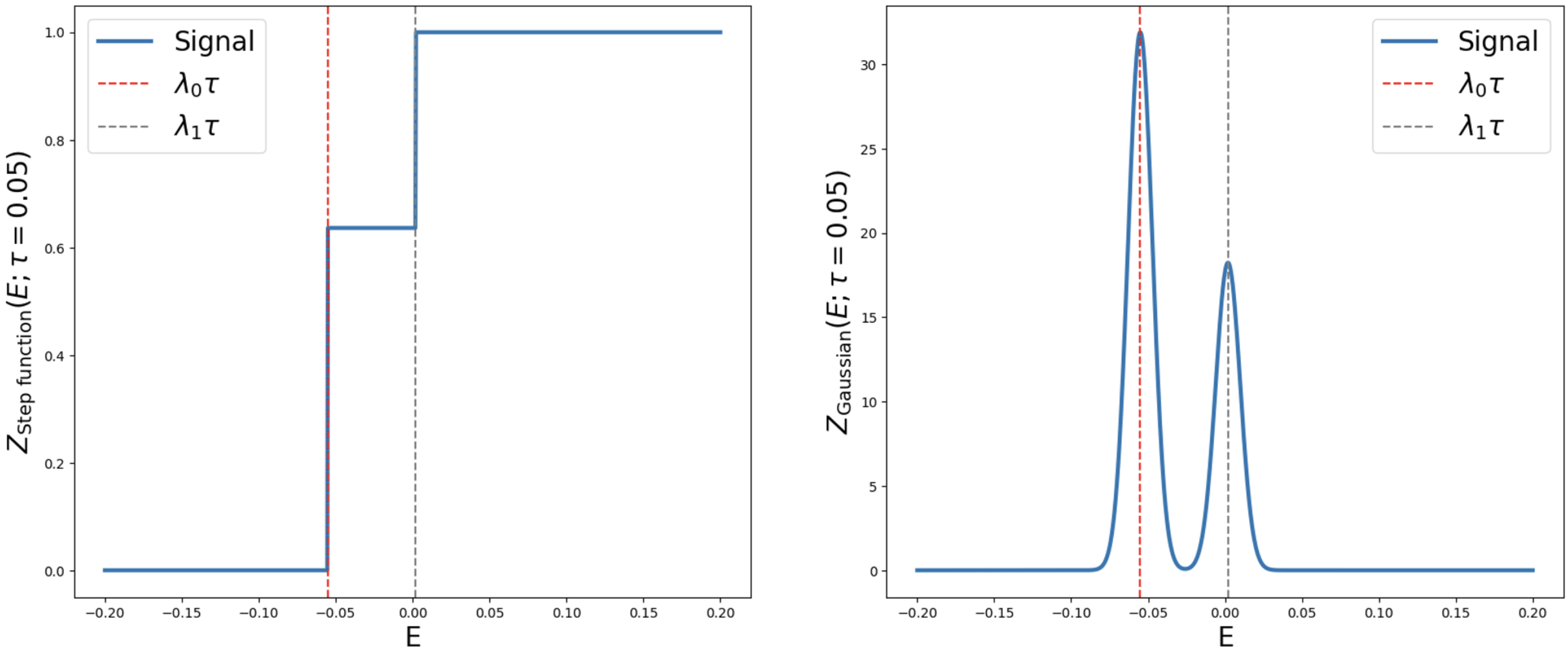}
    \caption{The ideal signal where there is no physical noise and statistical error, which is equivalent to $(F*\varrho)(x)$.}
    \label{fig:ideal_signal}
\end{figure}

\paragraph{Problem setup}
Given the Hamiltonian $H = \sum_{i} \lambda_i \ket{\lambda_i} \bra{\lambda_i}$, the algorithm starts with picking a state $|\psi\rangle$, whose overlap with the eigenstates $|\lambda_i\rangle$ of the Hamiltonian $H$ is $p_{i} = |\langle \lambda_i| \psi\rangle|^2$. This allows us to define a spectral density function
\begin{equation}
    \varrho(x) = \sum_{i} p_i\delta(x-\lambda_i\tau),
    \label{eq: spectral density}
\end{equation}
where $\delta(x)$ is the Dirac delta function. Then, we can pick a suitable convolution function $F(x)$ such that when the spectral density $\varrho(x)$ is convoluted by $F(x)$, the eigenvalues $\lambda_i$ are reflected as special points in the resulting function:
    \begin{equation}
        (F * \varrho)(x) = \int_{-T}^T F(y)\varrho(y-x)dy =\sum_{i}p_{i}F(x-\lambda_i \tau), \label{eq: convoluted_func}
    \end{equation}
where $T$ depends on the support of the function $F(x)$. The dimensionless parameter $\tau$ defines the search region of the algorithm, which is chosen to be a subset of $(-T, T)$ for all $i$. Its magnitude also affects energy resolution of the signal. The special points can be peaks, zeros, discontinuities, etc. For example, choosing $F(x)$ to be the periodic step function as in \cite{Lin:2021rwb, Kiss:2024sep, Wan:2021non, Blunt:2023gqs}:
\begin{equation}
    F(x) = \begin{cases}
        1, &\; x \in [2k\pi,\;(2k+1)\pi)\\
        0, &\; x \in [(2k-1)\pi,\;2k\pi)\\
    \end{cases},\quad
    k \in \mathbb{N} 
    \label{eq:peridoc_step}
\end{equation}
yields $(F*\varrho)(x)$ on the left hand side of Fig.~\ref{fig:ideal_signal}, where each jump corresponds to an eigenvalue. On the other hand, choosing $F(x)$ to be the Gaussian distribution as in \cite{Wang:2022gxu} with standard deviation $\sigma$:
\begin{equation}
    F(x) = \frac{1}{\sqrt{2\pi}\sigma}\exp\left({-\frac{x^2}{2\sigma^2}}\right)
\end{equation}
yields $(F*\varrho)(x)$ on the right hand side of Fig.~\ref{fig:ideal_signal}, where each peak corresponds to an eigenvalue. Next, the algorithm aims to quantum mechanically find a signal $Z_F(x;\tau)$ that mimics $(F*\varrho)(x)$.

\paragraph{Mimic $(F * \varrho)(x)$ quantum mechanically }
In Fourier space, $(F * \varrho)(x)$ takes the form 
\begin{equation}
(F * \varrho)(x)=\int_{-T}^{T}dk e^{ikx}\widetilde{F}(k)\sum_ip_i e^{-i k\lambda_i \tau}=\int_{-T}^{T}dk e^{ikx}\widetilde{F}(k)\langle e^{-ik\tau H}\rangle\approx \sum_{n=0}^{N-1}e^{ik_n x} \widetilde{F}(k_n) \langle e^{-ik_n\tau H}\rangle,
\label{eq: fouier_form}
\end{equation}
where $\langle e^{-i k_n \tau H}\rangle = \langle \psi|e^{-i k_n \tau H}|\psi\rangle$, $N$ is the number of Fourier modes considered, and 
\begin{equation}
    k_n = -T + \frac{2nT}{N}.
\end{equation}
 The integration upper bound $T = \pi$ if $F(x)$ is the periodic step function in eq.\ref{eq:peridoc_step}, and $T=\infty$ if $F(x)$ is chosen to be the Gaussian distribution.\footnote{In practice a cut-off is defined for $T$. For example in \cite{Wang:2022gxu} $T$ is cutoff at a value depending on the desired level of precision. The concrete cutoff value is reviewed at section \ref{subsubsec: benchmark param}.} We then arrive at the RHS of eq.~\ref{eq: fouier_form} by assuming that the Fourier transformed convolution function $\tilde{F}(k)$ can approximately be described as a sum over delta functions at discrete points $k_n$, where $n = 0, \cdots, N-1$. The significance of this is that it prescribes a procedure by which we can evaluate the convolution efficiently using a quantum processor. 

The expectation value $\langle e^{-iHk_n\tau}\rangle$ is what we will estimate with the Hadamard test in Fig.\ref{fig:hadamard_test}. The problem now becomes how many shots is supposed to be allocated to each $\langle e^{-iHk_n\tau}\rangle$ to keep the total run time close to the Heisenberg limit. To evaluate the convolution function $(F*\varrho)(x)$ via eq.~\eqref{eq: fouier_form} by sampling over $k_n$, so we rewrite the Fourier coefficient $\widetilde{F}(k_n)$ by:
\begin{equation}
    \widetilde{F}(k_n)=\mathcal{F}\frac{|\widetilde{F}(k_n)|}{\mathcal{F}}e^{i\varphi_n} \equiv \mathcal{F}\mathbb{P}(n)e^{i\varphi_n}, \; 0\leq n \leq N-1,
    \label{eq: fourier_coeff_rewrite}
    \end{equation}
where $\mathcal{F} = \sum_{n=0}^{N-1}|\widetilde{F}(k_n)|$, $\varphi_n = \arg \widetilde{F}(k_n)$, so that $\mathbb{P}(n)$ can be interpreted as a probability distribution function, as it is normalized to $1$. This yields
\begin{equation}
    (F*\varrho)(x)\approx Z_F(x;\tau)=\mathcal{F}\sum_{n=0}^{N-1}e^{ik_nx}e^{i\varphi_n}\mathbb{P}(n)\langle e^{-i k_n \tau H}\rangle,
    \label{eq: summand}
\end{equation}
which allows us to estimate in a Monte-Carlo manner by \emph{classically} sampling the probability distribution function $\mathbb{P}(n)$. Suppose the total number of samples we draw from $\mathbb{P}(n)$ is $N_{\text{sample}}$, and the number of samples being distributed to the Fourier mode $k_n$ is denoted as $N_n$, 
$2N_n$ number of shots will be allocated to estimate the expectation value $\langle e^{-ik_n \tau H}\rangle$, i.e.,
\begin{equation}
    \langle e^{-iHk_n\tau}\rangle\approx \frac{1}{N_n}\sum_{m=1}^{N_n}(X_m^{(n)}+iY_m^{(n)}),
    \label{eq: exp_sample_form}
\end{equation}
with $X_m^{(n)}+iY_m^{(n)} \in \{\pm1\pm i,\pm1\mp i\}$. Here, $X_m^{(n)}$ ($Y_m^{(n)}$) is defined as the $m$-th measurement result of the 0-th qubit from the Hadamard circuit in Fig.~\ref{fig:hadamard_test} with $t = k_n \tau$ and $i=0$ ($i=1$). Putting everything together, eq.\ref{eq: summand} can be evaluated using eq.\ref{eq: exp_sample_form} and becomes:

\begin{subequations}
\begin{align}
    Z_F(x;\tau)&\approx
    \frac{\mathcal{F}}{N_{\text{sample}}}\sum_{n=0}^{N-1}e^{ik_n x}e^{i\varphi_n} N_n\langle e^{-ik_n \tau H}\rangle \label{eq: signal_evo_rep}
    \\
    &\approx\frac{\mathcal{F}}{N_{\text{sample}}}\sum_{n=0}^{N-1}\sum_{m=1}^{N_n}e^{ik_n x}e^{i\varphi_n} (X_m^{(n)}+iY_m^{(n)}) \label{eq: signal_sample_rep}
    \end{align}
\end{subequations}
$Z_F(x;\tau)$ is the signal we aim to collect quantum mechanically and the rest of the job is to classically post-process the signal and read off special points from $Z_F(x;\tau)$. The whole procedure is summarized in Algorithm \ref{alg: SPE}.

\begin{algorithm}[t]
    \caption{Statistical Phase Estimation (SPE)}\label{alg: SPE}
    \SetKwComment{Comment}{$\triangleright$\ }{}
    \SetKwProg{myproc}{Procedure}{}{}
    \SetKwFunction{SigCol}{Signal Collection}
    \myproc{\SigCol{$H$, $N_{\text{sample}}$, $\widetilde{F}(k_n)$, $N$, $\tau$, $\{x_0, \cdots, x_{K-1}\}$}}{
        $\vec{Z} \gets \{Z_0, \cdots,\; Z_{K-1}\}\ = \{0, \; \cdots, \;0\}$\;
        $\mathcal{F} \gets \sum_{n}|\widetilde{F}(k_n)|$\;
        $\mathbb{P}(n) \gets \frac{|\widetilde{F}(k_n)|}{\mathcal{F}}$\;
        \For {$m \gets 0, \cdots, N_{\text{sample}}-1$}{
            $k_m \gets \text{Sample}(\mathbb{P}(n))$ \Comment*[r]{Classically sample from the probability distribution function $\mathbb{P}(n)$.}
            $\varphi_m \gets \arg\left(\widetilde{F}(k_m)\right)$\;
            $t_m \gets k_m \tau$\;
            $X_m \gets \text{HadamardTest}(H, |\psi\rangle, t_m, i=0)$ \Comment{Run Hadamard test in fig \ref{fig:hadamard_test} with $i = 0$}
            $Y_m \gets \text{HadamardTest}(H, |\psi\rangle, t_m, i=1)$ \Comment{Run Hadamard test in fig \ref{fig:hadamard_test} with $i = 1$}
            \For{$k = 0, \cdots, K-1$}{
                $Z_k \gets Z_k + \frac{\mathcal{F}}{N_{\text{sample}}}e^{i\varphi_m} e^{ik_m x_k}(X_m + i Y_m)$\;
            }
    }
    \Return $\vec{Z}$
}

\SetKwFunction{SPE}{SPE}
    \myproc{\SPE{$H$, $N_{\text{sample}}$, $\widetilde{F}(k_n)$, $N$, $\tau$}}{
        $\vec{x} \gets \{x_0, \cdots, x_{K-1}\}$ \Comment*[r]{Identify a search range. The range depends on the details of individual SPE algorithm.}
        $\vec{Z} \gets \textsc{Signal Collection}(N_{\text{sample}},\; \widetilde{F}(k_n),\; N,\; \tau,\; \vec{x})$\;
        Post process $\{\vec{x}, \vec{Z}\}$ to identify the position of the special feature $x_k^* \in \vec{x}$ that corresponds to the desired eigenvalue.\;
        \Return $x_k^*$
    }

\end{algorithm}

\subsubsection{Classical post-processing schemes}\label{subsec: post-proccessing}
After obtaining $Z_F(x)$, the last step of the algorithm is to process it and read off the desired eigenvalues. Here, we briefly review the post-processing schemes introduced in LT22 \cite{Lin:2021rwb} and Gaussian filter \cite{Wang:2022gxu}. We also introduce a slight modification to the Gaussian filter that can improve success probability.

\paragraph{LT22}
The LT22 algorithm uses the periodic step function as $F(x)$, where the first jump corresponds to the GS eigenvalue. Their classical post-processing scheme searches the GS energy by identifying the middle point $x_m$ of an interval $[L, R]$. Suppose the signal value evaluated at the middle point $Z_F(x_m; \tau)> \frac{3}{4}\eta$ , where $\eta$ is the prerequisite lower bound of $p_0$, we shrink the interval by fixing $L$ and decreasing $R$. Conversely, if $Z_F(x_m; \tau) < \frac{3}{4}\eta$, we shrink the interval by fixing $R$ and increasing the value of $L$. It is repeated until $x_m$ converges. In \cite{Lin:2021rwb}, every time we shrink the interval, $N_{\text{batch}}$ signals are collected and evaluated at $x_m$. Then, majority voting is used to determine if it is more likely if $Z_F(x_m;\tau) > \frac{3\eta}{4}$ or $Z_F(x_m;\tau) < \frac{3\eta}{4}$.

\paragraph{Gaussian Filter}
In the Gaussian Filter algorithm\cite{Wang:2022gxu}, instead of collecting a signal like the right hand side of Fig~\ref{fig:ideal_signal}, they aim to collect its derivative, i.e., the equation:
\begin{equation}
    Z'_F(x;\tau)\approx
    \frac{i\mathcal{F}}{N_{\text{sample}}}\sum_{n=0}^{N-1}k_ne^{ik_n x}e^{i\varphi_n} N_n\langle e^{-ik_n \tau H}\rangle
\end{equation}
is evaluated. Then, the eigenvalues become the zeros of the signal. The algorithm also requires a rough estimate to the GS eigenvalue $E_{\text{rough estimate}}$ beforehand that is $\frac{\sigma}{4}$-close to the exact GS energy, so that one can search in the interval

\begin{equation}
    x\in \left[E_{\text{rough estimate}} - \frac{\sigma}{4}, E_{\text{rough estimate}} +\frac{\sigma}{4}\right]
\end{equation}

\noindent to look for the zero. However, we find that a good rough estimate that satisfies this condition is generally not available and often requires a fine-tuned choice of $\sigma$ for the algorithm to succeed. Thus, in this article, we introduce an alternative method below.

\paragraph{Gaussian Fitting} As we know that the signal would locally look like 

\begin{equation}
Z=\frac{p_{0}}{\sqrt{2\pi}\sigma}\exp\left(-\frac{(x-\lambda_0 \tau)^2}{2\sigma^2}\right)    
\end{equation}

\noindent near $E_{\text{rough estimate}}$, we collect $M$ points of the signal in the interval 

\begin{equation}
    x\in \left[E_{\text{rough estimate}} - n_{\sigma} \frac{\sigma}{4}, E_{\text{rough estimate}} + n_{\sigma} \frac{\sigma}{4}\right].
    \label{eq: gaussian serach range}
\end{equation}

\noindent
Ideally, $n_{\sigma}$ should be $4\Delta_{\text{True}}/\sigma$ for the interval in eq.~\ref{eq: gaussian serach range} to include only the desired peak, where $\Delta_{\text{True}}$ is the exact energy gap between the ground state and first excited state. In practice, we don't know the value of $\Delta_{\text{True}}$, so we need to pick $n_{\sigma}$ depending on prerequisite knowledge of it. 

Then we define the cost function:

\begin{equation}
    L(P,\Lambda)=\frac{1}{M}\sum_{i=1}^M \left| Z_F(x_i,\tau) - \frac{P}{\sqrt{2\pi}\sigma} \exp{(-\frac{(x-\Lambda \tau)^2}{2\sigma^2}}) \right|^2
\end{equation}
\noindent
to be minimized. Then the optimal values $(P^*, \Lambda^*)=\argmin_{P,\Lambda} L(P,\Lambda)$, should correspond to the correct GS overlap $p_0$ and the GS eigenvalue $\lambda_0$,

\begin{equation}
(p_0,\lambda_0)=(P^*, \Lambda^*)=\argmin_{P,\Lambda} L(P,\Lambda).    
\end{equation}

\subsection{Key Indicators of algorithm performance}\label{subsec: spe indicators}

Here, we describe the performance indicators relevant to EFTQC algorithms. As mentioned above, the quantum part of the algorithm consists of a controlled time evolution operator as shown in Fig \ref{fig:hadamard_test}. Thus, at the logical level, the quantum resource consumption is approximately set by:
\begin{enumerate}
    \item Maximal evolution time $T_{\text{max}}$
    
    EFTQC algorithms trade evolution time for shorter circuit depth compared to FTQC, and the circuit depth is usually determined by the maximal evolution time required by the algorithm, while the total number of quantum circuit executions may be relatively high compared to an algorithm like QPE. For this reason, algorithms that can provide results with lower execution time require a smaller error correction overhead and can be carried out in the presence of more noise. The maximal evolution time is defined as

    \begin{equation}
    T_{\text{max}}=\max_i |t_i|    
    \end{equation}
    
    where $t_i$ is the evolution time of the $i$-th Hadamard test circuit call. While the controlled time evolution operator can be implemented by various means, and certain implementations allow trade-offs between circuit depth and qubit counts, at the algorithmic level $T_\text{max}$ is the limiting factor in terms of device capability.

    \item Total evolution time $T_{\text{total}}$
    
    The total evolution time is the sum of circuit execution times required to run the algorithm, disregarding classical resources required in the mean-time and parallelizability of the algorithm. It is defined as
    
    \begin{equation}
    T_{\text{total}} = \sum_{i}|t_i|.    
    \end{equation}

    As we shall see, the Gaussian filter implementation of statistical implementation defines a trade-off between this total execution time and the maximum circuit depth. This indicator is affected by the number of shots and individual circuit execution time and thus can be treated as the indicator of the total quantum resources the algorithm demands. We require that the total evolution time scales according to the Heisenberg limit:
    \begin{equation}
    T_{\text{total}} = \mathcal{O}(\epsilon^{-1})    
    \end{equation}
\end{enumerate}

In ground-state energy estimation problems, the complexity of the above indicators depend on quantities that characterize the concrete problem at hand. These quantities are:

\begin{itemize}
    \item Ground state overlap $(p_0)$
    
    Although we require our algorithm to estimate the ground state energy, while using only approximate ground states, EFTQC and FTQC algorithms generally are sensitive to the overlap of the input state with the actual ground state.

    \item Normalization factor $(\tau)$
    
    As the phase is a periodic quantity, it is desirable to normalize the spectrum into a suitable region. The $\tau$ parameter in eq.\ref{eq: spectral density} is responsible for controlling this normalization, and an ideal choice would be 
    
    \begin{equation}
    \tau=\frac{\pi}{4 \lVert H\rVert_2},    
    \end{equation}
    
    which guarantees the spectrum being normalized into the interval $[-\frac{\pi}{4}, \frac{\pi}{4}]$. However, this is not practical for large-sized Hamiltonian because computing the norm $\lVert H\rVert_2$ is effectively computing the maximal eigenvalue, which is classically impossible for large-sized Hamiltonian. On the other hand, we cannot choose $\tau$ to be arbitrarily large because that will cause the eigen-energies become too close to be distinguished.
    

    \item Energy gap between the ground state and the first excited state $(\Delta_{\text{True}})$:
    
    A narrow gap between the ground state and first excited state could potentially decrease the distinguishability of the ground and first excited state signal, in order to better resolve their energy difference, additional sampling using Hadamard tests is required in some instances.
    
    \item The precision we want to reach $(\epsilon)$
    
    Ideally, the total evolution time scales inversely with respect to $\epsilon$, i.e. the Heisenberg limit.
\end{itemize}

Given the discussions above, it is important to understand how the indicators $T_{\text{max}}$ and $T_{\text{total}}$ scale with respect to these quantities. In addition, the algorithms often require us to have the knowledge of the lower bounds of $p_0$ and $\Delta$ in order for us to pick parameters for classical post-processing. We hereby show the complexity of the $T_{\text{max}}$ and  $T_{\text{total}}$ as estimated by the papers.

\begin{table}[]
    \centering
    \begin{tabular}{|c|c|c|}
         \hline
        Algorithm & $T_{\text{max}}$ & $T_{\text{total}}$ \\
        \hline
        \hline
        LT22 & ${\mathcal{O}}(\epsilon^{-1} \text{polylog}(p_0^{-1}\tau^{-1}\epsilon^{-1}))$ & $\mathcal{O}(\epsilon^{-1} p_0^{-2}\text{polylog}(\epsilon^{-1}p_0^{-1} \tau^{-1}))$ \\
        \hline
        \hline
        Gaussian Filter & $\widetilde{\mathcal{O}}(\epsilon^{-\alpha} \Delta_{\text{True}}^{-1+\alpha})$ & 
$\widetilde{\mathcal{O}}(p_0^{-2}\epsilon^{-2+\alpha} \Delta_{\text{True}}^{1-\alpha})$
\\
\hline
    \end{tabular}
    \caption{Complexity of $T_{\text{max}}$, $T_{\text{total}}$ in terms of the energy accuracy $\epsilon$, normalization factor $\tau$ and lower bound of GS overlap $p_0$ and energy gap lower bound $\Delta_{\text{True}}$.}
    \label{tab:complexity}
\end{table}

Although LT22 and Gaussian Filter are variations on the same algorithm, they exhibit very different complexities, both in run-time and circuit depth. The most striking difference in these estimates is the dependence on the spectral gap, which is absent from LT22. The Gaussian filter algorithm allows a trade-off where $T_\text{total}$ may be increased in favor of reduced $T_\text{max}$, as expressed by the parameter $\alpha$ that lies in the interval $[0, 1]$. This increase in $T_\text{total}$ comes from the increased query complexity needed to calculate the filter function with a Gaussian filter that has a lower standard deviation. In return, it allows better $T_\text{max}$ scaling, which is important in the EFTQC era, where only partial error correction is available. On the contrary, since the filter function is a Gaussian, it is generally harder to resolve eigen-energies that are close together, which is why there is a strong dependence on the spectral gap both in  $T_\text{max}$ and $T_\text{total}$.

\section{Energy accuracy scaling of SPE}\label{sec: SPE simulation}
In section \ref{subsec: spe indicators}, the theoretical scaling of energy accuracy $\epsilon$ is reviewed. In this section, we explicitly demonstrate the scalings with actual simulation data. In this section, we perform our simulations with 2 settings. First, we perform the ideal simulation with classical estimation of $\langle e^{-iHk_n \tau}\rangle$ in  eq. \ref{eq: signal_evo_rep}. Then, we use eq. \ref{eq: signal_sample_rep} where $\{X_m^{(n)}, Y_m^{(n)}\}$ are sampled from a Hadamard test circuit whose controlled time evolution circuit is implemented with Trotterization. This introduces shot noise as well as Trotterization error to the simulation, so that we can study how Trotterization error pollutes the ideal scaling.

\subsection{Benchmark setup}\label{subsec: benchmark setting}
We start by explaining our benchmarking setups. We describe the system under consideration and the parameters that are used to run our simulations. Our computations here are carried out by the open-source quantum algorithm development library QURI Parts \cite{quri_parts}.

\subsubsection{Target system and input state}
Here we use \ce{H2} with the \ce{H} atoms \qty{1.0}{\angstrom} apart as the target system. The input state in the full CI basis with sto-3g is 
\begin{equation}
    |\psi\rangle=\cos\theta|0011\rangle+\sin\theta|0110\rangle,
\end{equation}
where $\ket{0011}$ is the Hartree-Fock state and $\theta$ is chosen such that the GS overlap is about 77\%. The Hartree-Fock state with $\theta=0$ will give us a GS overlap over 90\% and should be able to yield better results. However, we deliberately choose a slightly-lower-GS-overlap input state to show that the algorithms can still perform well when the state preparation is not ideal.

\subsubsection{Benchmark parameter choices}\label{subsubsec: benchmark param}
\paragraph{LT22:} The number of Fourier modes is given by $2d+1$, where we scan over $d=50, 100,\allowbreak 500, 1000,\cdots, 500000, 1000000$. The number of samples required scales as \cite{Lin:2021rwb}:
\begin{equation}
    N_{\text{sample}} = \mathcal{O}\left(\eta^{-2} \log^2(d)\left(\log\log(\epsilon^{-1}\tau^{-1})+\log\vartheta\right)\right)
    \label{eq: LT22_N_sample_complexity}
\end{equation}
where $1-\vartheta$ is the success probability. With $\vartheta$ chosen to be $10^{-12}$, $\tau=\frac{1}{20}$, $\eta = 0.5p_0$, the maximal number of samples is approximately 10000, which we choose to fix to throughout later benchmarks. Finally, recall that in section \ref{subsec: post-proccessing}, multiple signals are generated in the INVERT CDF sub-procedure, here in our benchmarks, we only generate 1 signal and use the CERTIFY sub-procedure to identify the jump.

\paragraph{Gaussian:} The parameter choice for Gaussian filter requires prior knowledge of the lower bound of $p_0$, $\Delta_{\text{True}}$ and a very rough estimation of the eigenvalue. The rough estimation is given by the output of LT22 with total shots of 10000 and $d=100$. Here we denote the lower bound of $p_0$ as $\eta$ and lower bound of $\Delta_{\text{True}}$ as $\Delta$. The Fourier transform is evaluated by discretizing to 1000 Fourier modes and the integration region is limited in the interval $[-T, T]$. Here, we follow \cite{Wang:2022gxu} where $T$ and the standard deviation $\sigma$ relate to the desired accuracy $\epsilon$ through
\begin{gather}
    \sigma=\min \left(\frac{0.9\Delta}{\sqrt{2\ln(9\Delta \epsilon^{-1} \eta^{-1})}}, 0.2\Delta\right) \\
    T=\pi^{-1}\sigma^{-1}\sqrt{2 \ln(8\pi^{-1}\tilde{\epsilon}^{-2}\eta^{-1})}, \; \tilde{\epsilon}=\frac{0.1 \epsilon \eta}{\sqrt{2\pi}\sigma^3}.
\end{gather}
In the subsequent benchmarking results, we choose $\Delta=\Delta_{\text{True}}^{4/5}$, $\eta=0.7p_0$ and scan over $\epsilon=[10^{-2}, 10^{-3}, \cdots,10^{-6}]$. The number of $x$ we evaluate $Z_F(x;\tau)$ on, $M$,  is suggested to be:
\begin{equation}
    M=\Big\lceil{\frac{\sigma}{\epsilon}\Big\rceil}+1
\end{equation}
$M$ is usually too small to be useful for further post-processing when $\epsilon > 10^{-3}$. So, we set the lower bound of $M$ to be at least 1000. The number of samples is suggested to be
\begin{equation}
    N_{\text{sample}}=\Bigg\lceil\frac{\mathcal{F}^2\ln\frac{4M}{\delta}}{\epsilon^2}\Bigg\rceil,
\end{equation}
where $\delta$ is the probability of the initial guess falling into the $\frac{\sigma}{4}$-close region. 

\subsection{Ideal simulation of statistical phase estimation}

Having established the benchmark setting, we first start with an ideal simulation where the expectation value $\langle e^{-iHt}\rangle$ in \ref{eq: signal_evo_rep} is classically computed with matrix multiplication.  With the settings in the last subsection, we present our simulation results below. We will plot the energy error $\Delta E = |E_{\text{exact}} - E_{\text{phase estimation}}|$ against the performance indicators $T_{\text{max}}$ and $T_{\text{total}}$.
\begin{figure}
    \centering
    \includegraphics[width=1\linewidth]{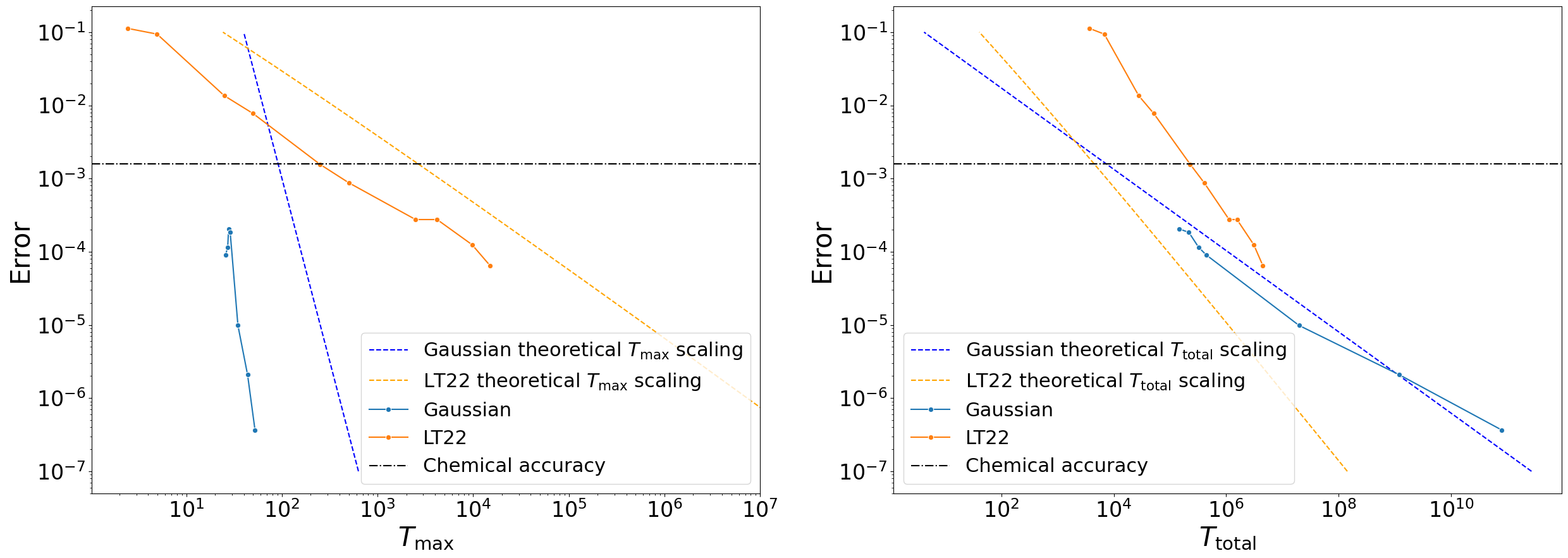}
    \caption{
    Ideal simulation result where the expectation value $\langle e^{-iHt}\rangle$ is computed exactly via matrix multiplication. The dashed lines are drawn with the complexities in table~\ref{tab:complexity}. It is shown that our simulation result is consistent with the scaling of the theoretical complexity estimation.
    }
    \label{fig:benchmark}
\end{figure}


The dashed line on Fig.~\ref{fig:benchmark} indicates the theoretical estimated $T_{\text{max}}$ and $T_{\text{total}}$ complexity, which were shown in table\ref{tab:complexity}. Our numerical experiment shows scaling consistent with the theoretical estimations. Note that the LT22 $T_{\text{total}}$ dashed line scales as the Heisenberg limit’s scaling, up to $\text{polylog}$ factors. This is the line we wish the algorithms to adhere to as closely as possible. On the other hand, the LT22 $T_{\text{max}}$ dashed line is the line we wish our algorithm to improve from in order to decrease circuit depth. The dramatic difference in $T_{\text{max}}$ is due to the nature of the Fourier transform of the convolution function, which we discuss in the next paragraph.

In our benchmarks, we have aimed to show the different strengths of LT22 and Gaussian filter. In particular we kept $T_{\text{max}}$ relatively constant for Gaussian filter, while in our simulations $T_{\text{total}}$ was increased to obtain a lower error. This is in contrast to LT22 where both $T_{\text{max}}$ and $T_{\text{total}}$ scale with the error, but the scaling of $T_{\text{total}}$ is somewhat worse for the Gaussian filter method. It is worth understanding the $T_{\text{max}}$ behavior. To do so, we take our attention back to the first line of eq.\ref{eq: signal_evo_rep}. It shows that the signal is the expectation value $\langle e^{-ik\tau H}\rangle$ weighted by the probability distribution $\mathbb{P(n)}$. This suggests that the maximum evolution time $T_{\text{max}}$ of which  $\langle e^{-i k_n\tau H}\rangle$ is estimated is determined by $\mathbb{P}(n)$. We can then plot the evolution time $t=k_n\tau$ against the probability $\mathbb{P}(n) = \mathbb{P}\left(\frac{t}{\tau}\right)$ it is drawn from.

\begin{figure}
    \centering
    \includegraphics[width=0.8\linewidth, height=0.48\linewidth]{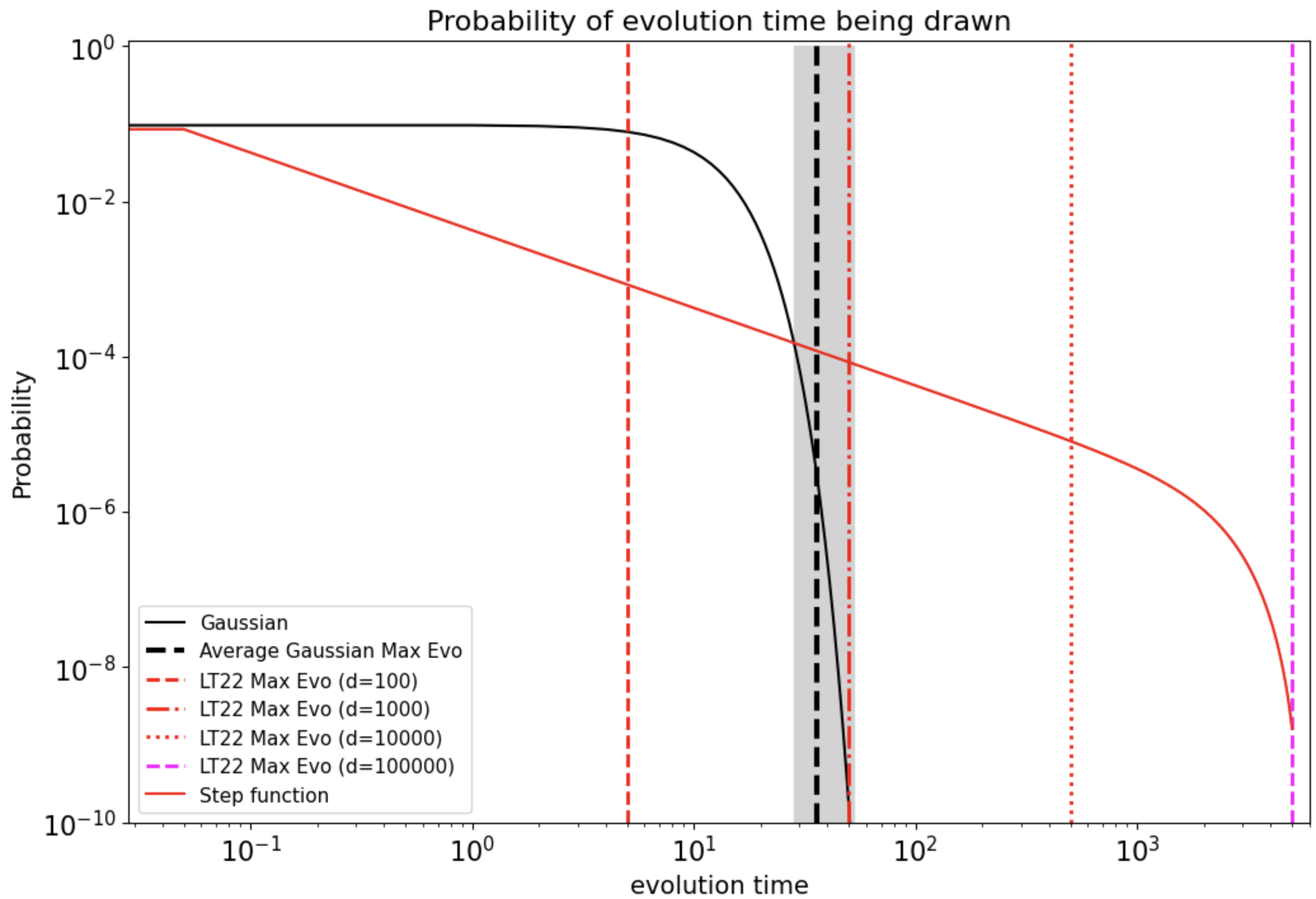}
    \caption{The probability of an evolution time being drawn. The gray band corresponds to the range of $T_{\text{max}}$ of the 7 runs of the Gaussian filter algorithm presented in Fig~\ref{fig:benchmark}.}
    \label{fig:prob_evo}
\end{figure}

The Gaussian filter method, in contrast to LT22, does not define a strict cutoff for the evolution time. Instead the shot count required for circuit evaluation is distributed over controlled time-evolution circuits with the distribution defined by the Fourier transform of the Gaussian filter itself. The black line in Fig.~\ref{fig:time_evo_t50} shows the resulting probability distribution. To avoid cluttering, the gray band corresponds to the range of the $T_{\text{max}}$ values over 7 Gaussian filter runs, with the number of samples ranging from $2\times 10^4$ to $10^{10}$. The vertical black line corresponds to their average. We can see that due to the exponential decaying behavior, the average $T_{\text{max}}$ is much less than those LT22 runs whose $d > 1000$. This explains why the $T_{\text{max}}$ value barely scales with respect to the energy accuracy $\epsilon$ in the Gaussian filter algorithm.

\subsection{Encoding the time-evolution operator using the Trotter-Suzuki decomposition}\label{subsec: trotter benchmark}
As there is no known circuit implementation that allows us to obtain the exact $\langle e^{-iHt}\rangle$ with only 1 access to the quantum circuit, we use the Hadamard test circuit whose controlled-time evolution is implemented based on Trotter decomposition:
\begin{equation}
    e^{-iHt} \approx \left[\prod_{i=1}^N e^{-i \frac{h_i t}{r} P_i}\right]^r,\quad
    H = \sum_{i=1}^{N} h_i P_i
\end{equation}
Here, we benchmark with 50 Trotter steps, i.e. $r=50$.

\begin{figure}[t!]
    \centering
    \subfigure[
        Simulation with $\langle e^{-iHt}\rangle$ evaluated by a Hadamard test where the controlled time evolution is implemented by Trotterization. Here, the Trotter step is set to be 50. The ideal line is plotted against those performed with Trotter decomposition.
    ]{\includegraphics[width=1\linewidth]{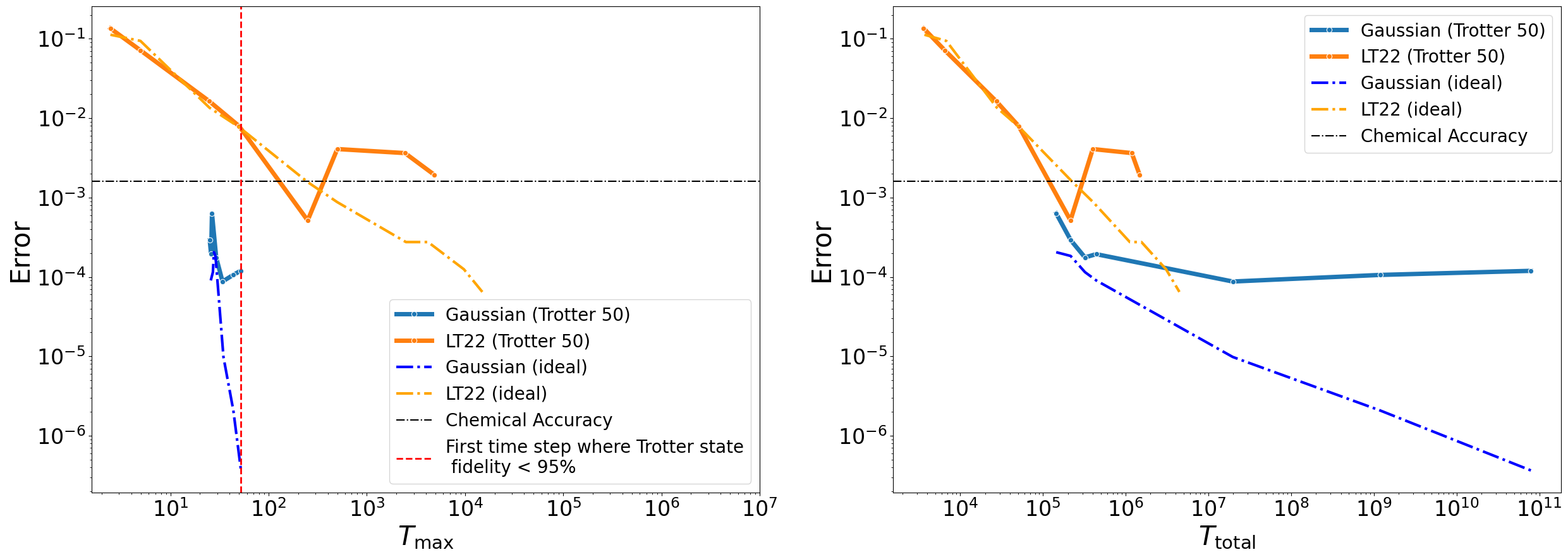}}
    \subfigure[Time dependence of state fidelity with 50 Trotter steps.]{\includegraphics[width=0.7\linewidth, height=0.4\linewidth]{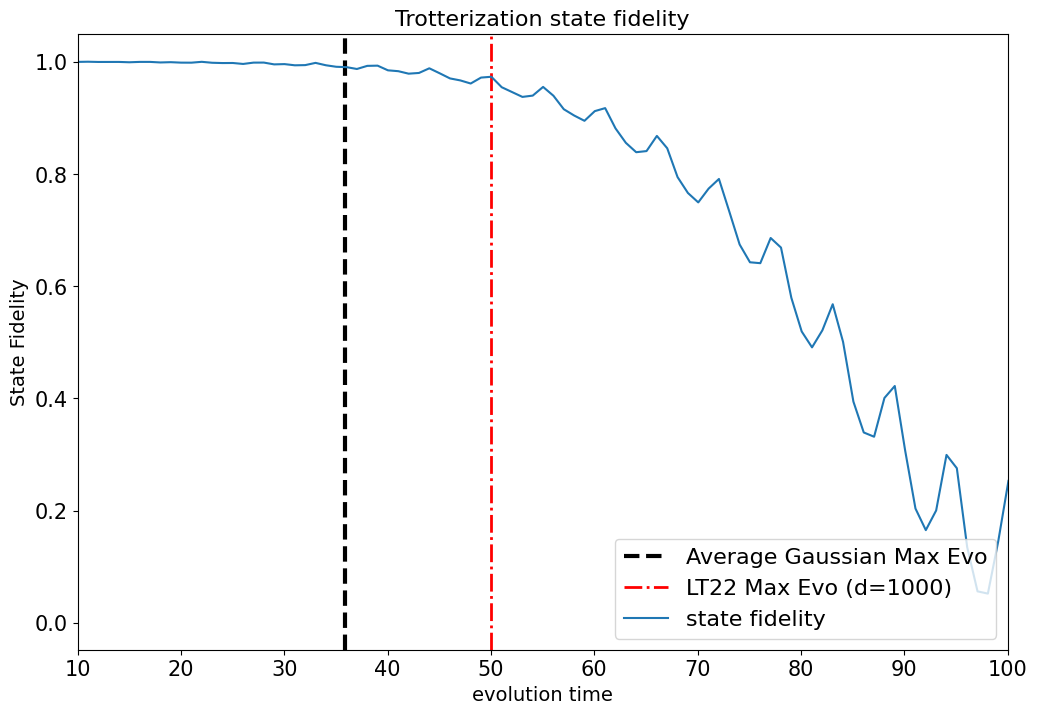}}
    
    \caption{Simulation with a 50-Trotter-step circuit with the corresponding Trotter state fidelity.}
    \label{fig:time_evo_t50}
\end{figure}

Fig.~\ref{fig:time_evo_t50} shows how Trotterization error corrupts the algorithm performance. As $T_{\text{max}}$ grows, Trotterization error for long time evolution degrades the estimation of $\langle e^{-iHt}\rangle$.  The red dashed line represents the point where the fidelity of the Trotterized state fall beneath 95 \%.  As a clearer illustration, we look at the Trotterization time evolution state fidelity compared to the exact evolution state in figure~\ref{fig:time_evo_t50}(b).

As we can see from Fig.\ref{fig:time_evo_t50}(b), while the state fidelity remains very close to 1 until 50 time steps, the Gaussian line is still affected by Trotterization error even if the state fidelity is larger than 95 \%. For LT22, we can see that the Trotter line starts to deviate from the ideal line after the vertical red dashed line in Fig~\ref{fig:prob_evo}. This is approximately where the state fidelity enters the rapid oscillating region in Fig~\ref{fig:time_evo_t50}, causing the performance to be unpredictable. This reinforces the requirement that EFTQC algorithm should aim for decreasing $T_{\text{max}}$ as much as possible.

\section{Benchmarking statistical phase estimation against VQE}\label{sec: compare with VQE}
The last section contrasts the performance of LT22 and Gaussian filter SPE against VQE. We first study the shot counts consumed by SPE and VQE on noiseless devices. Next, we add physical noise to our simulation with our previously outlined noise model derived from a simplified view of the STAR architecture (see Section~\ref{sec:star_architecture}). This leads to our major result figure \ref{fig:noisy_shots_benchmarks}, which demonstrates that with proper error correction architecture, performance of SPE becomes comparable to VQE when the physical error rate is at the order of $10^{-4}$. Once the physical error rate decreases below $10^{-4}$, the performance of SPE has significant improvement over VQE in both energy accuracy and real device execution time. 

\subsection{Comparison with VQE on noiseless device}\label{subsec: noiseless shot compare}

\begin{figure}
    \centering
    \includegraphics[width=1\linewidth, height=0.5\linewidth]{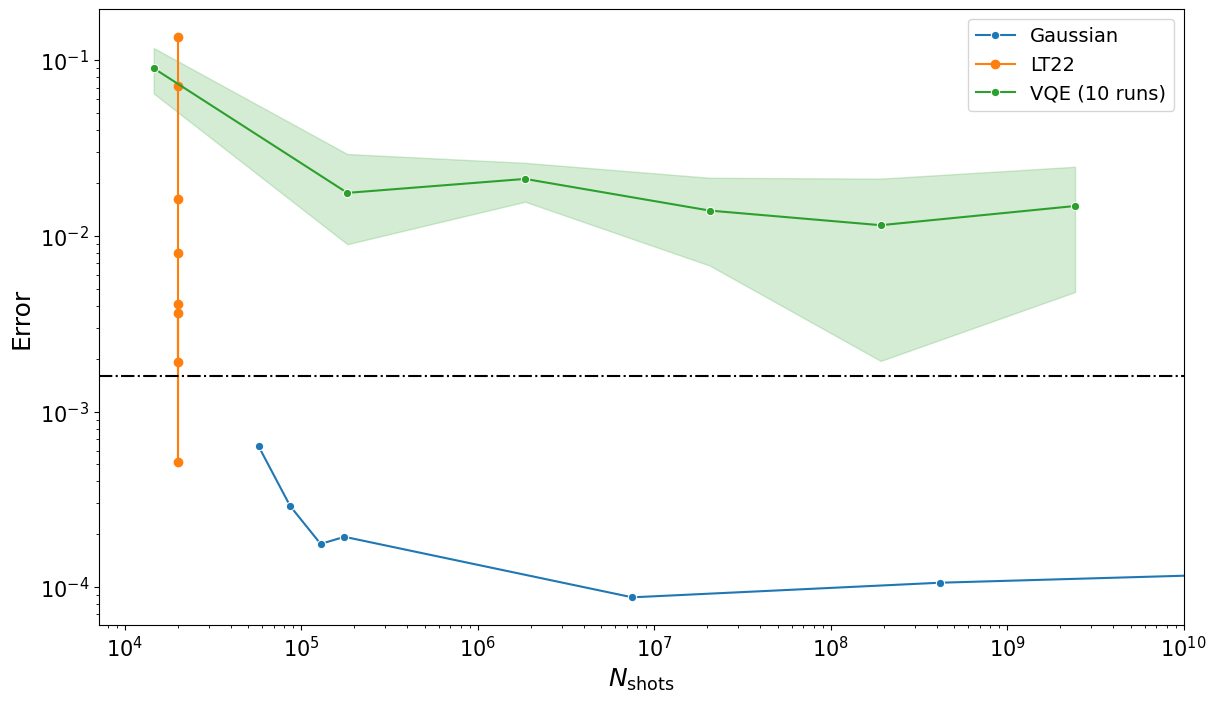}
    \caption{Comparison of VQE against various EFTQC algorithms. Error is defined as the energy difference between the result energy and FCI energy. $N_{\text{shots}}$ is the number of shots used to obtain an estimated energy. VQE is executed 10 times with the 1-uCJ ansatz. As mentioned above, the parameter used for running LT22 is chosen to be $10^4$ for all choices of $d$, thus the line in the plot is vertical. For VQE, $N_{\text{shots}}$ includes all shots used to optimize circuit parameters.}
    \label{fig:vqe_comparison}
\end{figure}

We first quantitatively show that the SPE can outperform VQE with fewer shots with figure~\ref{fig:vqe_comparison}. We use the same benchmarking setup as in section \ref{subsec: benchmark setting}. In addition, VQE is performed with the 1-uCJ ansatz with a varied number of shots and the BFGS algorithm as the optimizer. We compare the VQE energy error against that of LT22 and Gaussian filter. From the LT22 line, we see that it reaches better energy accuracy with fewer shots required. This is due to the nice scaling of the shot complexity in Eq.~\ref{eq: LT22_N_sample_complexity}. From the Gaussian line, the original paper estimates the number of shots scales as $\epsilon^{-2}$, which is approximately the same as the VQE scaling. This behavior is demonstrated in Fig.~\ref{fig:vqe_comparison} as well. However, with the same number of shots, the Gaussian filter can achieve much lower energy error.

 
\subsection{Realizing quantum advantage in the EFTQC era}
 Finally, we address the question: \emph{At what physical error rate will statistical phase estimation be advantageous compared to VQE on real devices?}
We investigate the relation between energy error against shot counts and real device execution time under different physical error rates.

The real device execution time is estimated assuming the device runs on the STAR architecture reviewed in Section~\ref{sec:star_architecture}. To estimate the execution time, our logical level circuit is compiled down to architecture-level instructions, each with their own instruction execution time and fidelity. The circuit execution time is then estimated by summing over the execution time of each instruction. As the instruction success fidelity is also estimable, we can estimate the overall logical circuit fidelity by multiplying the instruction fidelity, which is shown in table \ref{tab:Circuit Fidelity}. 
This serves as an architecture-independent indicator of how imperfect a given logical circuit is, and Fig.~\ref{fig:noisy_shots_benchmarks} allows us to understand how resilient each algorithm is under imperfect circuit execution.

\subsubsection{Benchmark settings for noisy simulation}
Now, we modify our benchmark settings based on the results in section \ref{subsec: trotter benchmark}. As we have already experimented with Trotter error, for our results elucidate on the consequence of physical noise, we
\begin{enumerate}
    \item Isolate the pollution due to physical noise from that of the Trotter error to better understand the effect due to imperfect devices.
    \item Avoid unnecessarily long real device execution time and shot counts.
\end{enumerate}
 Based on these requirements, we make the following modifications.
\begin{enumerate}
    \item For LT22, as we see in figure \ref{fig:time_evo_t50}(b), the Trotter error comes in to pollute the result approximately from $d=1000$. We run the algorithm with $d=1000$, but with $N_{\text{sample}} = 10^4,\;10^5,\;10^6,\; 10^7$ so that we can see if longer execution time can give us better results.
    \item For Gaussian filter, the Trotter error is reflected by a worse $T_{\text{total}}$ scaling, which is illustrated in the right hand side of figure \ref{fig:time_evo_t50} (a). To tame the bad scaling, we set the maximal number of shots to be $10^7$.
\end{enumerate}

\subsubsection{Algorithm performance in the presence of varying noise intensities}\label{subsubsec: varying_noise}
\begin{figure}
     \centering
     \includegraphics[width=1\linewidth]{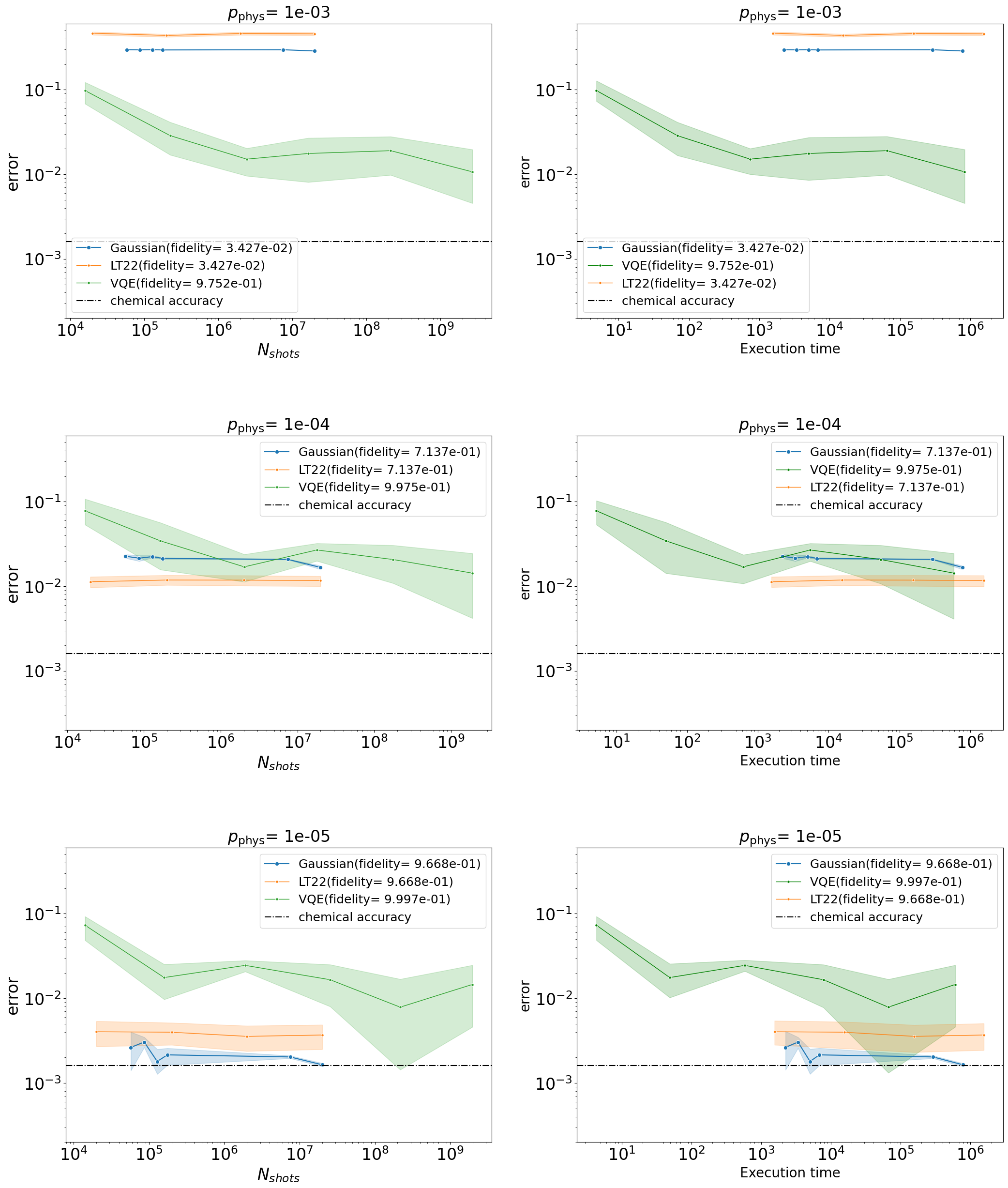}
     \caption{
     Energy error against shot count and estimated real device execution time. The execution time is in unit of seconds.
     }
     \label{fig:noisy_shots_benchmarks}
 \end{figure}

Here, we run the algorithms with $p_{\text{phys}}= 10^{-3}, \; 10^{-4}, \;10^{-5}$.  Following the section \ref{subsec: noiseless shot compare}, we first investigate how energy error scales with $N_{\text{sample}}$ under noisy setting. As seen on Fig.~\ref{fig:noisy_shots_benchmarks}, the simulation result demonstrates that the energy error of statistical phase estimation is comparable to VQE starting from $p_{\text{phys}} \leq 10^{-4}$. The circuit fidelities shown in the plot legend indicate that once the circuit fidelity exceeds 70\%, the energy estimation by SPE surpasses that of VQE at  $> 97\%$ circuit fidelity. This is a concrete evidence of the noise resilience nature of EFTQC algorithm as alluded to at the bottom of section~\ref{subsec: algo comparison}.

\begin{table}[]
    \centering
    \begin{tabular}{|c|c|c|c|}
        \hline
        Algorithm & LT22 & Gaussian & VQE \\
        \hline
        \hline
        Fidelity($p_{phys} = 10^{-3}$) & $3.427 \%$ & $3.427 \%$ & $97.52 \%$\\ 
        \hline
        Fidelity($p_{phys} = 10^{-4}$) & $71.37 \%$ & $71.37 \%$ & $99.75 \%$\\ 
        \hline
        Fidelity($p_{phys} = 10^{-5}$) & $96.68 \%$ & $96.68 \%$ & $99.97 \%$\\ 
        \hline
    \end{tabular}
    \caption{Circuit fidelity of different algorithms under different physical error rates. The LT22 and Gaussian algorithm are executed on 50-Trotter-stepped circuits of different rotation angles. VQE is executed with the 1-uCJ ansatz.}
    \label{tab:Circuit Fidelity}
\end{table}
 
\begin{figure}
     \centering
     \includegraphics[width=1\linewidth]{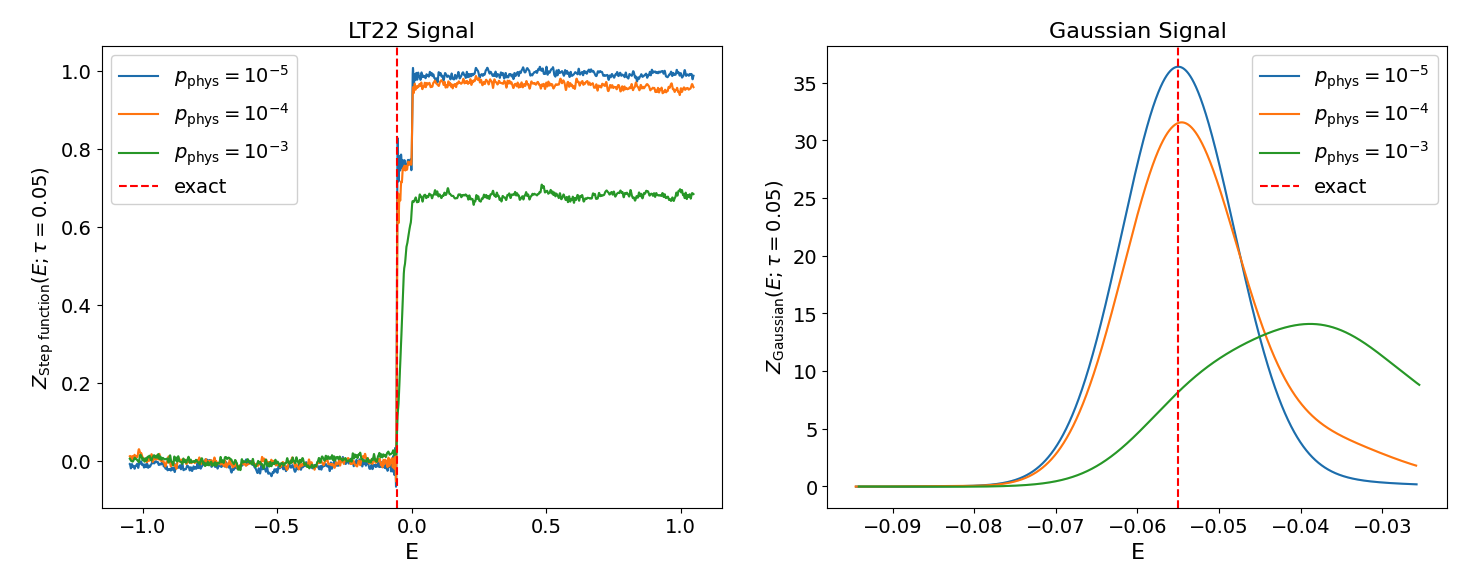}
     \caption{Signal Collected from noisy simulation. We scan over $p_{phys} = 10^{-3}, \;10^{-4}, \; 10^{-5}$ for both LT22 and Gaussian filter. Starting from $p_{phys} \leq 10^{-4}$, the first jump in the LT22 signal and the peak in the Gaussian signal are close to the exact value.}
     \label{fig:noisy_signal}
 \end{figure}

\paragraph{Noisy signal and robustness under noise}
To get a clearer idea of the effects of noise on the statistical phase estimation, we plot out the collected signal before the classical post-processing under different $p_{\text{phys}}$ settings. As seen in Fig.~\ref{fig:noisy_signal} there is a stark difference between LT22 and Gaussian filter when it comes to the impact of noise on the signal.

When the physical error rate is less than $10^{-4}$, we find that the discontinuity of the LT22 signal is very close to the ideal signal and for Gaussian filter, the peak is very close to the exact energy. However, neither the peak nor the discontinuity are sharp enough to eventually fall into the chemical accuracy regime. 

\subsubsection{Benchmarking the energy error against algorithm execution time}
Finally, we return to the main topic of our study, i.e. understanding the algorithm execution time needed to reach the desired level of accuracy. In our settings, the logical circuit for all samples is the same up to the rotation angles of the $R_Z$ gates, reflecting different evolution times. Thus, each shot will take the same amount of time to be executed.

The benchmark results on the right-hand side column of figure \ref{fig:noisy_shots_benchmarks} show that phase estimation is a lot slower than VQE. This is due to the fact that the Hadamard test circuit is much deeper than the 1-uCJ circuit. However, we point out two avenues of further plausible improvements.


\begin{enumerate}
    \item In our simulations, we use 50-Trotter-step time evolution circuits consistently for all samples. We have not scaled the number of Trotter steps with the time evolution required for each sample individually. Alternatively, there are time evolution compilations that can be done to reduce circuit depth, which are not part of this study, but are expected to reduce the required runtime \cite{Mizuta:2022rrz, khatri2019quantum, cirstoiu2020variational, Commeau:2020aab}. 
    \item In addition, as statistical phase estimation does not involve optimization loops, it is possible to perform the Hadamard tests in parallel to reduce execution time. However, this is limited by the number of logical qubits and quantum devices available to us.
\end{enumerate}

In summary, from our noisy simulation result we are able to identify the level of physical error rate and circuit fidelity required to make SPE favorable over VQE. We point out that we reached this advantage without relying overly much on fine-tuning of parameters or making significant modifications to SPE. However, looking for instance at Fig~\ref{fig:noisy_shots_benchmarks}, we see that even with a relatively noisy device we are able to collect a signal which has a clearly visible inflection point at the ground state energy. It seems that clever signal processing has the potential to allow for more accurate determination of ground state energies than we were able to achieve with LT22, even for very noisy devices. As a direct example of recent developments in this direction, we refer to \cite{Kiss:2024sep}.


\section{Challenges and future prospects}\label{sec: prospect}
SPE and its variants have positioned themselves to play a pivotal role in the coming era of partially error-corrected quantum devices. Throughout the previous sections we have elaborated on how these algorithms function and benchmarked their performance. We first investigated the run-time scaling with energy accuracy both in terms of maximal and total circuit execution time for the \ce{H2} molecule. Then we went on to compare SPE with VQE. Our findings suggest using SPE outperforms VQE when the physical error rate is low enough.

While we await the arrival of devices that possess error correction architectures there are still important challenges to address which may greatly improve the performance of these algorithms. One such challenge is reducing parameter choice. Achieving chemical accuracy with EFTQC algorithms often requires delicate tuning of post-processing parameters. This tuning necessitates an understanding of the relation between the system characteristics, e.g. $\eta$ and $\Delta$ used throughout in section \ref{subsubsec: benchmark param}, and the choice of algorithm hyperparameters. From the user perspective, a good quantum algorithm should not be that dependent on user expertise outside of the necessary domain knowledge that relates to the problem at hand. Instead, the algorithm itself should rely on as few hyperparameters as possible and if possible be robust to bad parameter choice to some extent. For examples of algorithms which were developed specifically to reduce the need for parameter fine-tuning, we refer the reader to QMEGS \cite{Ding:2024qvu} and RUPTURES \cite{Kiss:2024sep}.

Another significant challenge is improving time evolution circuit compilers. For a Trotterized circuit to accurately simulate the time evolution operator, the circuit depth must increase significantly, which can introduce physical errors and degrade the performance of the algorithm. This highlights the need for more advanced circuit compilers such as LVQC\cite{Mizuta:2022rrz}, qDrift \cite{PhysRevLett.123.070503, Chen:2020bga}, qSwift \cite{Nakaji:2023gze}, Variational Fast Forwarding \cite{cirstoiu2020variational, Commeau:2020aab}, Qubitization \cite{Low:2016znh} and various other methods \cite{Mansuroglu:2023oji, Bilek:2022cpo, khatri2019quantum}. Research into EFTQC phase estimation algorithms and quantum simulation is essential, as these areas can synergistically enhance each other to facilitate near-term practical applications.

Furthermore, once we have a good compiler for time evolution, we need to develop a specialized transpiler to optimize the circuits further. This transpiler should be customized for each specific hardware and architecture to reduce errors.

Although we've only touched on a few challenges here, they highlight the great potential of EFTQC research to create new opportunities across various fields, with promising developments likely soon.


\begin{acknowledgements}
    We would like to thank Yuya O. Nakagawa, Shoichiro Tsutsui and Masaya Kohda for proof-reading and commenting on a draft version of this paper. We also thank Toru Kawakubo for discussions on the noise model for the STAR architecture, Shoichiro Tsutsui for comments on the Trotterization error, Ryota Kojima and Toru Shibamiya for discussions on the problem formulations.
\end{acknowledgements}

\newpage

\bibliography{mybib}

\end{document}